\documentclass[sigconf,screen]{acmart}
\usepackage{multirow}
\usepackage{subfigure}
\usepackage{blindtext}
\usepackage{enumitem}
\usepackage{algorithm}  
\usepackage{algpseudocode}  
\usepackage{amsmath}  
\usepackage{hyperref}
\usepackage{flushend}
\usepackage{balance}
\usepackage{mathtools}
\usepackage{bbm}

\AtBeginDocument{%
\providecommand\BibTeX{{%
  \normalfont B\kern-0.5em{\scshape i\kern-0.25em b}\kern-0.8em\TeX}}}

\copyrightyear{2023} 
\acmYear{2023} 
\setcopyright{acmcopyright}
\acmConference[KDD '23]{the 29th ACM SIGKDD Conference on Knowledge Discovery and Data Mining}{August 6--10, 2023}{Long Beach, CA, USA}
\acmBooktitle{the 29th ACM SIGKDD Conference on Knowledge Discovery and Data Mining (KDD '23), August 6--10, 2023, Long Beach, CA, USA}
\acmPrice{15.00}
\acmDOI{10.1145/xxxxxx.xxxxxx}
\acmISBN{978-1-4503-xxxx-x/xx/xx}

\begin{document}

\title{PIER: Permutation-Level Interest-Based End-to-End Re-ranking Framework in E-commerce}

\author{Xiaowen Shi}
\authornote{Equal contribution. Listing order is random.}
\affiliation{%
 \institution{Meituan}
 \city{Beijing}
 \country{China}
}
\email{shixiaowen03@meituan.com}

\author{Fan Yang}
\authornotemark[1]
\affiliation{%
 \institution{Meituan}
 \city{Beijing}
 \country{China}
}
\email{yangfan129@meituan.com}

\author{Ze Wang}
\authornote{Corresponding author.}
\affiliation{%
 \institution{Meituan}
 \city{Beijing}
 \country{China}
}
\email{wangze18@meituan.com}

\author{Xiaoxu Wu}
\affiliation{%
 \institution{Meituan}
 \city{Beijing}
 \country{China}
}
\email{wuxiaoxu04@meituan.com}

\author{Muzhi Guan}
\affiliation{%
 \institution{Meituan}
 \city{Beijing}
 \country{China}
}
\email{guanmuzhi@meituan.com}

\author{Guogang Liao}
\affiliation{%
 \institution{Meituan}
 \city{Beijing}
 \country{China}
}
\email{liaoguogang@meituan.com}

\author{Yongkang Wang}
\affiliation{%
 \institution{Meituan}
 \city{Beijing}
 \country{China}
}
\email{wangyongkang03@meituan.com}

\author{Xingxing Wang}
\affiliation{%
 \institution{Meituan}
 \city{Beijing}
 \country{China}
}
\email{wangxingxing04@meituan.com}

\author{Dong Wang}
\affiliation{%
 \institution{Meituan}
 \city{Beijing}
 \country{China}
}
\email{wangdong07@meituan.com}
\renewcommand{\shortauthors}{Xiaowen Shi and Fan Yang, et al.}

\begin{abstract}
  % \renewcommand{\thefootnote}{\fnsymbol{footnote}}
  % \footnotetext[2]{Equal contribution. Listing order is random.}
  % \renewcommand{\thefootnote}{\fnsymbol{footnote}}
  % \footnotetext[3]{This work was done when Chuhang Zhang was an intern in Meituan.}
  % \renewcommand{\thefootnote}{\fnsymbol{footnote}}
  % \footnotetext[1]{Corresponding author.}
  % \renewcommand{\thefootnote}{\fnsymbol{footnote}}
  
 Re-ranking draws increased attention on both academics and industries, which rearranges the ranking list by modeling the mutual influence among items to better meet users' demands. Many existing re-ranking methods directly take the initial ranking list as input, and generate the optimal permutation through a well-designed context-wise model, which brings the evaluation-before-reranking problem. Meanwhile, evaluating all candidate permutations brings unacceptable computational costs in practice. Thus, to better balance efficiency and effectiveness, online systems usually use a two-stage architecture which uses some heuristic methods such as beam-search to generate a suitable amount of candidate permutations firstly, which are then fed into the evaluation model to get the optimal permutation. However, existing methods in both stages can be improved through the following aspects. As for generation stage, heuristic methods only use point-wise prediction scores and lack an effective judgment. As for evaluation stage, most existing context-wise evaluation models only consider the item context and lack more fine-grained feature context modeling.

 This paper presents a novel end-to-end re-ranking framework named \textit{PIER} to tackle the above challenges which still follows the two-stage architecture and contains two mainly modules named FPSM and OCPM. Inspired by long-time user behavior modeling methods, we apply SimHash in FPSM to select top-K candidates from the full permutation based on
user’s permutation-level interest in an efficient way. Then 
we design a novel omnidirectional attention mechanism in OCPM to better capture the context information in the permutation. Finally, we jointly train these two modules in an end-to-end way by introducing a comparative learning loss, which use the predict value of OCPM to guide the FPSM to generate better permutations. Offline experiment results demonstrate that PIER outperforms baseline models on both public and industrial datasets, and we have successfully deployed PIER on Meituan food delivery platform.
  
\end{abstract}

% <concept_id>10002951.10003227.10003447</concept_id>
% <concept_desc>Information systems~Computational advertising</concept_desc>
% <concept_id>10002951.10003260.10003272</concept_id>
% <concept_desc>Information systems~Online advertising</concept_desc>
% <concept_id>10002951.10003260.10003282.10003550</concept_id>
% <concept_desc>Information systems~Electronic commerce</concept_desc>

% \ccsdesc[500]{Information systems~Computational advertising}
% \ccsdesc[500]{Information systems~Online advertising}
% \ccsdesc[500]{Information systems~Electronic commerce}

\maketitle

\section{Introduction}

\begin{figure}[tb]
  \centering
  \includegraphics[width=1.0\linewidth]{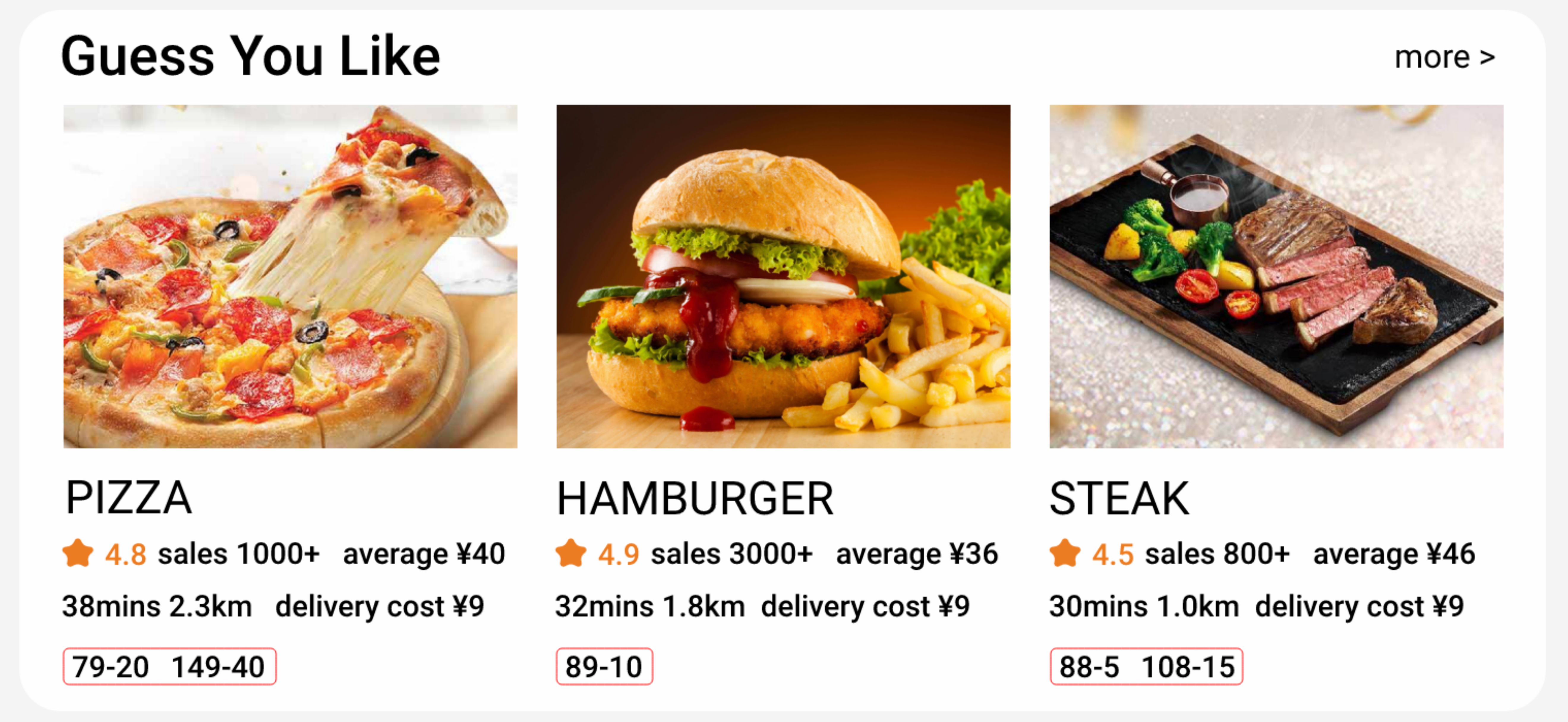}
  \caption{Guess You Like in Meituan.}
  \label{fig:guess}
\end{figure}

\label{sec:1}

E-commerce applications such as JD.com and Meituan have a large number of items. To improve the user's decision-making efficiency, a slate which contains limited items is usually provided based on the user's interest. As shown in figure \ref{fig:guess}, a slate named Guess You Like which consists of three items, is displayed to users on Meituan food delivery platform. 

Due to the rapid growth of deep learning techniques, many well-designed ranking models have been proposed to improve the recommendation performance, mainly focusing on feature interaction (e.g. Wide\&Deep \cite{cheng2016wide}, DeepFM \cite{guo2017deepfm}, xDeepFM \cite{lian2018xdeepfm}), user interest modeling (e.g. DIN \cite{zhou2018DIN}, DIEN \cite{zhou2019dien}, ETA \cite{chen2021eta}), and so on. However, most existing ranking methods only model the CTR of the current item, but ignore the crucial mutual influence among contextual items. In order to model the influence of the arrangement of displayed items on user behaviors, the re-ranking stage is introduced to rearrange the initial list from the ranking stage. 
\begin{sloppypar}
Existing re-ranking methods can be divided into two categories \cite{feng2021grn}. The first category is the step-greedy re-ranking strategy  \cite{bello2018seq2slate,zhuang2018midnn,gong2022real,feng2021grn}, which sequentially decides the display results of each position. Such methods only consider the preceding information but ignore the succeeding information, which is insufficient to obtain the optimal result. Different from the greedy strategy, another solution is the context-wise re-ranking strategy \cite{pei2019prm,ai2018dlcm,xi2021crum,chen2022extr,feng2021PRS}, which uses a context-wise evaluation model to capture the mutual influence among items and re-predict the CTR of each item. Methods like PRM \cite{pei2019prm}, directly take the initial ranking list as input, and generates the optimal permutation based on the predict value given by context-wise model. These one-stage methods suffer an evaluation-before-reranking problem \cite{xi2021crum}, that is, due to the order obtained after re-ranking is different from the initial order, inputting the re-ranked list will result in different prediction results. In order to resolve evaluation-before-reranking problem, a straightforward solution is to evaluate every possible permutation, which is global-optimal but is too complex to meet the strict inference time constraint in industrial system. Therefore, most existing re-ranking framework uses a two-stage architecture \cite{xi2021crum,feng2021PRS} which consists of permutation generation and permutation evaluation. To be specific, they use some heuristic methods such as beam-search \cite{reddy1977beamsearch} to generate the suitable amount of candidate permutations firstly and then feed into the evaluation model to get the optimal permutation. 
\end{sloppypar}
In order to improve the performance of the re-ranking stage under the two-stage architecture, on the one hand, the generated candidate permutations should contain the optimal permutation as much as possible. On the other hand, the context-wise evaluation model should predict as accurately as possible. However, existing methods in both stages can be improved through the following aspects:

\begin{itemize}[leftmargin=*]
\item Generation stage. Some heuristic methods, such as the beam search algorithm \cite{feng2021PRS}, merely use point-wise prediction scores (i.e. item CTR) to generate candidate permutations, while ignoring the mutual influence between each item and its contexts in one permutation. In addition, since the generation stage is independent of the evaluation stage, the evaluation results cannot guide the generating process. Therefore, the quality of the generated candidate permutations lacks an effective judgment.

\item Evaluation stage. Different from the point-wise item prediction, the evaluation of the permutations needs to use various types of contextual information to fully model the mutual influence among items. In addition to the influence of item context, there is also the fine-grained influence of feature context. These features form a variety of channels, and users may be interested in the features of a certain channel. For example, price-sensitive users will pay more attention to the comparison of price information in the context, which we call it as multi-feature channel competition problem.

\end{itemize}

To resolve the aforementioned issues, we propose a novel end-to-end re-ranking framework named Permutation-Level Interest-Based End-to-End Re-ranking (PIER). Our framework still follows the two-stage paradigm which contains two mainly modules named Fine-grained Permutation Selection Module (FPSM) and Omnidirectional Context-aware Prediction Module (OCPM). Inspired by long-time user behavior modeling methods \cite{chen2021eta,cao2022sdim}, we apply SimHash in FPSM to select top-K candidates from the full permutation based on user's permutation-level interest in an efficient way. Then in OCPM, we design a novel omnidirectional attention and context-aware predict mechanism to better capture the context information and predict the list-wise CTR of each item in the permutation. Finally, we integrate these two modules into one framework and training in an end-to-end way. We introduce a comparative learning loss, which use the predict value of OCPM to guide the FPSM to generate better permutations.

The main contributions of our work are summarized as follows:

\begin{itemize}[leftmargin=*]
\item We propose a novel re-ranking framework named PIER, which integrates generation module and evaluation module into one model and can be trained in an end-to-end manner.
\item We conduct extensive offline experiments on both public dataset and real-world industrial dataset from Meituan. Experimental results demonstrate the effectiveness of PIER. It is notable that PIER has been deployed in Meituan food delivery platform and has achieved significant improvement under various metrics.
\end{itemize}

 \section{related work}

An industrial recommender system typically consists of three stages\footnote{We do not discuss advertising-related stages (e.g. mix-ranking \cite{liao2022crossdqn, chen2022hierarchically}, auctions \cite{liao2022nma}) in this paper.}: matching \cite{zhu2018tdm}, ranking \cite{wang2020cold, li2023decision} and re-ranking \cite{pei2019prm}. Matching stage aims to recall thousands of relevant items from the whole item set. Ranking stage \cite{cheng2016wide, guo2017deepfm, zhou2018DIN} point-wisely predict the click-through rate (or conversion rate, etc.) of recalled items. Re-ranking stage aims to find the best (e.g. maximization of the total clicks) permutation from the initial list given by the ranking model. In this paper, we mainly focus on re-ranking stage.  

Typical re-ranking methods can be divided into two categories. The first category is the step-greedy re-ranking strategy \cite{bello2018seq2slate,zhuang2018midnn,gong2022real,feng2021grn}, which sequentially decides the display results of
each position through recurrent neural network or approximation solution. Seq2slate \cite{bello2018seq2slate} utilizes pointer-network and MIRNN utilizes \cite{zhuang2018midnn} GRU to determine the item order one-by-one. Similarly, the client-side short video re-ranking framework of Kuaishou \cite{gong2022real} combines a point-wise prediction model which takes ordered candidates list as input with beam-search to sequentially generate the final list. These methods ignore succeeding information, which is insufficient to obtain the optimal result.

Another category is context-wise re-ranking strategy \cite{pei2019prm,ai2018dlcm,xi2021crum,chen2022extr,feng2021PRS}, which uses
a context-wise evaluation model to capture the mutual influence
among items and re-predict the CTR/CVR of each item. Methods such as PRM \cite{pei2019prm} and DLCM \cite{ai2018dlcm} take the initial ranking list as input, use RNN or self-attention to model the context-wise signal and output the predict value of each item. The optimal permutation is sorted according to the predict value. Such methods bring an evaluation-before-reranking problem \cite{xi2021crum} and leads to sub-optimum. Similarly, methods such as EXTR \cite{chen2022extr} estimate pCTR of each candidate item on each candidate position, which are substantially point-wise models and thus limited in extracting exact context. In order to model the exact context of the permutation, a straightforward solution
is to evaluate every possible permutation through a well-designed context-wise model. This is a global-optimal method
but is too complex to meet the strict inference time constraint in
industrial system. In order to reduce the complexity, PRS \cite{feng2021PRS} adopts beam-search to generate few candidate permutations firstly, and score each permutation through a permutation-wise ranking model. Although heuristic methods are effective, they only use point-wise prediction scores and lack an effective judgment, which can be improved further.

In our scenario, we still adopt the two-stage architecture and focus on improving the overall performance by optimizing both two stages under the online time-consuming constraints.

 \section{Preliminaries}
% wz-todo 下方cite里需加引文
Typically, an industrial recommender system consists of three consecutive stages: matching, ranking and re-ranking \cite{zhu2018tdm,li2023decision,wang2020cold,feng2021PRS,xi2021crum,pei2019prm}. Given a certain user involving his/her input ranking list $\mathcal{C}= \{a_i\}_{i=1}^{N_o}$, the final displayed $N_d$ items to user are formulated as the displayed list $\mathcal{P}= \{a_i\}_{i=1}^{N_d}$, where $N_d \leq N_{o}$. 

If there is no re-ranking stages, the top-$N_d$ items are selected as $\mathcal{G}$ from $\mathcal{C}$ for display.
Therefore, the task of re-ranking stage is to learning a re-ranking strategy $\pi: \mathcal{C} \rightarrow \mathcal{P}^{*}$, which aims to select and rearrange items from $\mathcal{C}$, and subsequently recommends a better displayed list $\mathcal{P^*}$, with the aim of improving indicators such as CTR and GMV.

\section{Methodology}
\begin{figure}[b]
  \centering
  \includegraphics[width=1\linewidth]{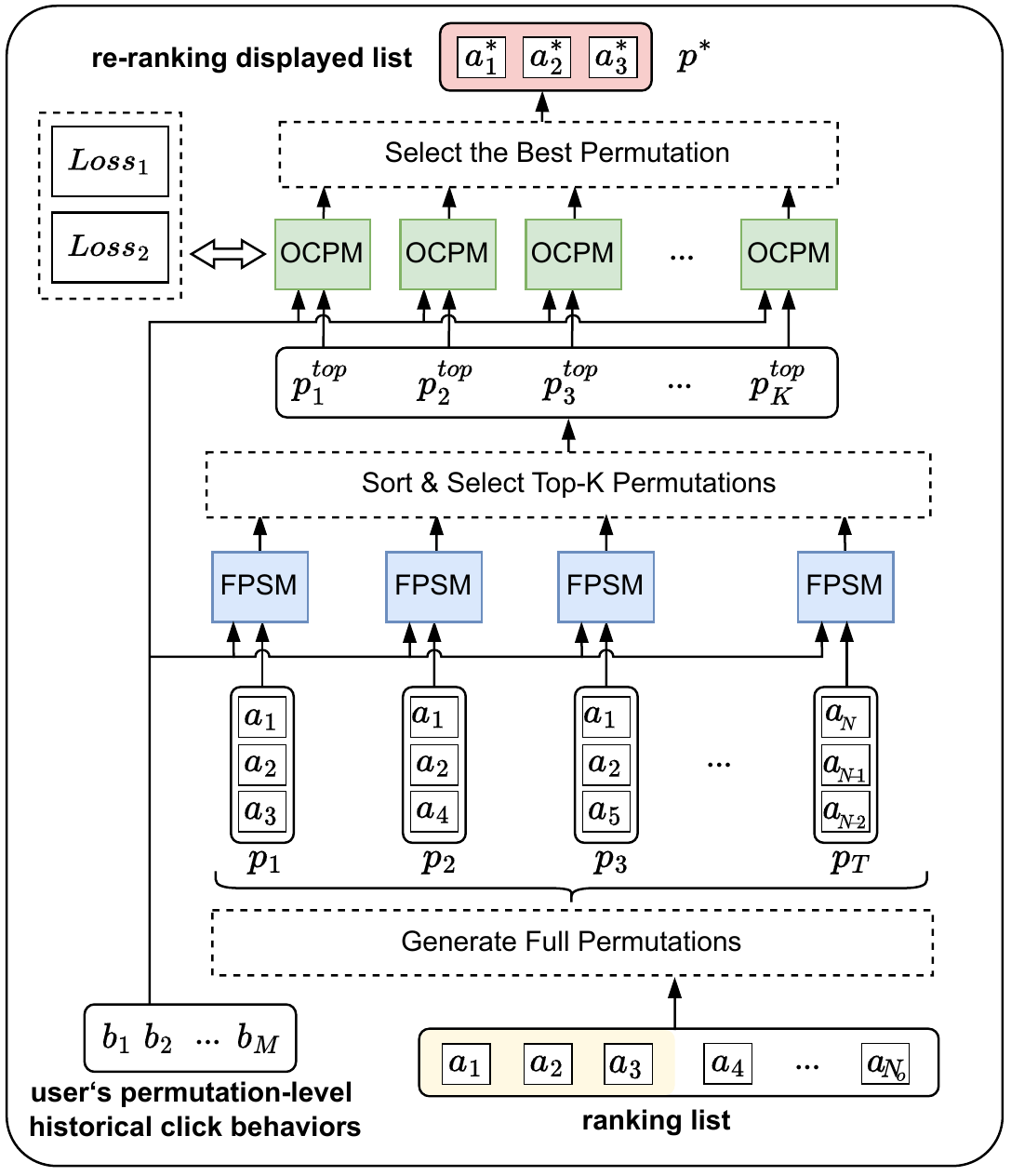}

  \caption{Overview of our framework PIER. PIER takes ranking list and user's permutation-level historical click behaviors as input, and outputs a re-ranking list to display with the help of FPSM and OCPM.}
  \label{fig:dca2}
\end{figure}

\begin{figure*}[tb]
  \centering
  \includegraphics[width=\textwidth]{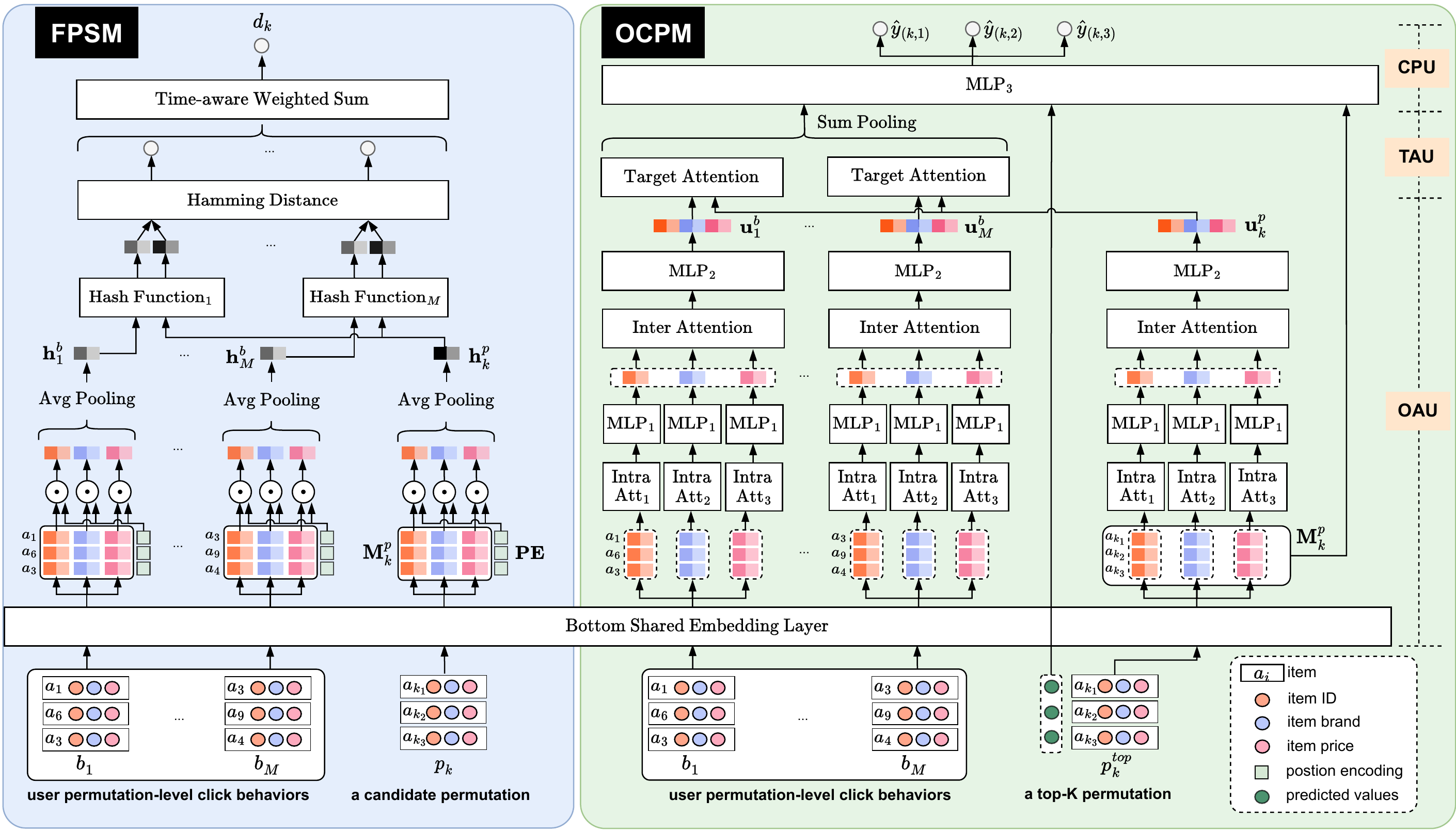}

  \caption{The structures of Fine-grained Permutation Selection Module (FPSM) and Omnidirectional Context-aware Prediction Module (OCPM). FPSM takes user permutation-level click behaviors and a candidate permutation as input and outputs the distance score of this permutation. OCPM takes user permutation-level click behaviors and a top-$K$ permutation selected by FPSM as input and outputs the predicted list-wise pCTR of each item in this permutation. Best view in color.}
  \label{fig:dca}
\end{figure*}

We present the overview structure of PIER (\textbf{P}ermutation-level \textbf{I}nterest-based \textbf{E}nd-to-End \textbf{R}e-ranking) in Figure \ref{fig:dca2}.
Specifically, we take ranking list $\mathcal{C}= \{a_i\}_{i=1}^{N_o}$ and user's permutation-level click behavior sequence $\mathcal{B}= \{b_i\}_{i=1}^{M}$ as the input of PIER. 
Then, we generate candidate permutations $\mathcal{G}= \{p_i\}_{i=1}^{T}$ on $\mathcal{C}$ through full permutation algorithm\footnote{We set the number of items $N_d$ in each permutation to 3 for illustration in figures.}. 
Next, we use the Fine-grained Permutation Selection Module (FPSM) to select top-$K$ permutations from large number of candidate permutations. 
Finally, we use the Omnidirectional Context-aware Prediction Module (OCPM) to calculate scores of each permutation and select the best permutation $p^*$ as the re-ranking list to display.

% wz-todo 下方cite里需加引文
In FPSM, we propose the time-aware hamming distance calculated based on SimHash \cite{chen2021eta,charikar2002simhash,DND4WC}. The top-K permutations are selected by sorting the distance  between user's permutation-level click behaviors and candidate permutations.
In OCPM, we design a novel omnidirectional attention unit
to model the context information in each permutation and output the list-wise pCTR of each item in this permutation. The best permutation is selected based on the output score, i.e., the sum of list-wise pCTRs.
The relationship between FPSM and OCPM is like the matching and ranking stages in recommender systems. We combine the two into one framework to generate the optimal re-ranking list.
Next we will detail FPSM and OCPM separately.

\subsection{Fine-grained Permutation Selection Module}
\label{sec:fpsm}
% wz-todo 下方cite里需加引文
For time-consuming considerations, some re-ranking methods utilize heuristic methods such as beam-search  \cite{feng2021PRS} to generate candidate permutations and use a well-designed prediction model to select the optimal permutation, but these heuristic methods are not consistent with the modeling target, leading to suboptimal performance. Inspired by long-term user behavior modeling methods such as ETA and SDIM  \cite{cao2022sdim}, we propose FPSM to select top-K candidates through SimHash. Here, We use the user's historical click behaviors as  target and then calculate the distance between it and all candidate permutations. If the distance is closer, we believe that the permutation can better match user's interest thus can bring greater revenue. In this manner, we can not only reduce the time complexity, but also can make consistent selection by training the FPSM and the prediction model in an end-to-end way. Next, we will introduce how to select the top-K permutations through FPSM.

As shown in the left part of Figure \ref{fig:dca}, we first use the bottom shared embedding layers to extract the embeddings from raw inputs. For permutation $p_k$, we denote the embedding matrix $\mathbf{M}^{p}_{k}$ as follows:

\begin{equation}
  \begin{aligned}
    \mathbf{M}^{p}_{k} & = 
    \left[ 
      \begin{array}{c}
        \mathbf{E}^{p_k}_{0_{ };_{{ }_{ }}} \\
        \mathbf{E}^{p_k}_{1_{ };_{{ }_{ }}} \\
        \dots \\
        \mathbf{E}^{p_k}_{N_d;_{{ }_{ }}} \\
    \end{array}
    \right] 
    \\
    & =  
    \left[ 
      \begin{array}{ccccc}
        \mathbf{E}^{p_k}_{{ };0} & \mathbf{E}^{p_k}_{{ };1}& \dots & \mathbf{E}^{p_k}_{{ };N_f}
    \end{array}
    \right] 
    \\
    & =  
    \left[
    \begin{array}{ccccc}
        \mathbf{e}^{p_k}_{0;0} & \mathbf{e}^{p_k}_{0;1}& \dots & \mathbf{e}^{p_k}_{0;N_f} \\
        \mathbf{e}^{p_k}_{1;0} & \mathbf{e}^{p_k}_{1;1}& \dots & \mathbf{e}^{p_k}_{1;N_f} \\
        \dots & \dots & \dots & \dots \\
        \mathbf{e}^{p_k}_{N_d;0} & \mathbf{e}^{p_k}_{N_d;1}& \dots & \mathbf{e}^{p_k}_{N_d;N_f} \\
    \end{array}
    \right] 
    \in \mathbb{R}^{N_d \times N_f \times D}, 
   \end{aligned}
\end{equation} 
where $N_d$ is the number of items in each permutation, $N_f$ is the number of feature fields (i.g., ID, category, brand and so on) in each item, $D$ is the dimension of the embedding-transformed feature field, 
$\mathbf{e}^{p_k}_{i;j} \in \mathbb{R}^{D}$ is the embedding of the $i$-th item's $j$-th feature field in permutation $p_k$,
$\mathbf{E}^{p_k}_{i;} \in \mathbb{R}^{N_f \times D}$ is the embedding matrix of the $i$-th item in permutation $p_k$, 
and $\mathbf{E}^{p_k}_{;j} \in \mathbb{R}^{N_d \times D}$ is the embedding matrix of the $j$-th feature field in permutation $p_k$.
Analogously, the embedding matrix of the $m$-th permutations in user's permutation-level history click behaviors is formulated as $\mathbf{M}^{b}_{m} \in \mathbb{R}^{N_d \times N_f \times D}$.

Next, we generate the position encoding matrix $\mathbf{PE} \in \mathbb{R}^{N_d \times D}$ for each permutation as follows:
\begin{equation}
  \begin{aligned}
    &\ \ \ \ \ \ \ \ \ \ \ \ \ \ \ \ \ \text{PE}_{(i,2d)} = \sin(i/10000^{2d/D}) ,\\
    &\ \ \ \ \ \ \ \ \ \ \ \ \ \ \text{PE}_{(i,2d+1)} = \cos(i/10000^{2d/D}) ,\\
    \mathbf{PE} = & \ 
    \left[ 
      \begin{array}{cccc}
        \text{PE}_{(0,0)} & \text{PE}_{(0,1)} &\dots& \text{PE}_{(0,D)} \\
        \text{PE}_{(1,0)} & \text{PE}_{(1,1)} &\dots& \text{PE}_{(1,D)} \\
        \dots&\dots&\dots&\dots \\
        \text{PE}_{(N_d ,0)} & \text{PE}_{(N_d ,1)} &\dots& \text{PE}_{(N_d ,D)} \\
    \end{array}
    \right]   \in \mathbb{R}^{N_d  \times D}.
  \end{aligned}
\end{equation}

Then, the embedding matrices of each feature field are multiplied by the position encoding matrix $\mathbf{PE}$ respectively and then merged into corresponding permutation representation $\mathbf{h}^{p}_k$ by average pooling, as follows:
\begin{equation}
  \begin{aligned}
    \mathbf{h}^{p}_k = \frac{1}{N_f} \sum_{i=1}^{N_f} \text{Avg-Pool} \Big(\mathbf{E}^{p_k}_{;i} \!\!\odot\! \mathbf{PE}
    \Big),  \ \  \forall k \in [N_o].
  \end{aligned}
\end{equation}

Analogously, the representation of $m$-th permutation in user's permutation-level history click behaviors is formulated as $\mathbf{h}^{b}_m$.

In our scenario, users are more likely to click the permutations which are closer to the user's interest. 
We use user's permutation-level history click behaviors to represent user's interest and calculate the distance between user's interest and each candidate permutation.
Specifically, we utilize the random projection schema (SimHash)  \cite{chen2021eta,charikar2002simhash,DND4WC} to calculate the similarity between the representations of user clicked permutations and the representations of candidate permutations. We first generate $M$ different hash functions  corresponding to $M$ user's permutation-level behaviors. For each candidate permutation $p_k$, we hash its representations $\mathbf{h}^{p}_k$ with $M$ different hash functions and calculate the hamming distance between it and each user's permutation-level behavior, as follows: 
\begin{equation}
  \begin{aligned}
   \text{Sim}(p_k,b_m) = \text{Hash}_{m} (\mathbf{h}^{p}_k, \mathbf{h}^{b}_m), \ \  \forall m \in [M],  \forall k \in [N_o].
  \end{aligned}
\end{equation}  

Meanwhile, the more recent behavior, the more it can reflect the user's current interest and will be given a higher weight in similarity calculation. So we weight these distance according to the occurrence time of each behavior to obtain the time-aware hamming distance, as follows: 
\begin{equation}
  \begin{aligned}
    d_{k} = \sum_{m=1}^Mw_{m} \cdot \text{Sim}(p_k,b_m),  \ \  \forall k \in [N_o].
  \end{aligned}
\end{equation}
where $w_{m}$ is the time-aware weight of $m$-th behavior.

Finally, we sort candidate permutations based on their distance scores and select top-$K$ permutations $\mathcal{P}^{top-K}= \{p_i^{top}\}_{i=1}^{K}$ with the smallest distance scores as the output of FPSM. 

Since FPSM shares bottom embedding layers with OCPM and fixes random vectors for hash function and position encoding, it dose not have its own independent parameters that need to be trained.
In order to ensure the quality of the selected top-K permutations during training, we propose a contrastive loss to improve the performance of FPSM. The detail of contrastive loss is discussed in Section \ref{join_train}.

\subsection{Omnidirectional Context-aware
Prediction Module}

For each candidate permutation selected by FPSM, we then use OCPM to predict the pCTR of each item through three consecutive unit: Omnidirectional Attention Unit (OAU), Target Attention Unit (TAU), Context-aware Prediction Unit (CPU). The architecture of OCPM is shown in the right part of Figure \ref{fig:dca} and we will introduce each unit of OCPM separately.

\subsubsection{Omnidirectional Attention Unit (OAU)}

When showing a permutation to users, on the one hand, they may pay attention to different displayed information such as price, delivery fee, rating, etc. On the other hand, the competitive relationship between the same feature of different items will also affect the user’s behavior. For instance, placing expensive items ahead cheap items can stimulate user's desire to click on cheap one. From this point of view, in OCPM, we design an omnidirectional attention unit to effectively modeling the information of each permutation.

Specifically, OCPM first uses the same bottom shared embedding layers as FPSM to extract embeddings from raw inputs. Then we use $N_f$ parameter-independent self-attention layers  \cite{vaswani2017attention} to calculate the mutual influence of different items in each field separately and output corresponding matrix $\mathbf{H}^{p_k}_{j}$, as follows:
\begin{equation}
  \begin{aligned}
    \mathbf{H}^{p_k}_{j} = \text{soft} \max (\frac{\mathbf{Q}^{p_k}_j{\mathbf{K}^{p_k}_j}^\top}{\sqrt{D}})\mathbf{V}^{p_k}_j, \ \ \forall j \in [N_f], \forall k \in [K], 
  \end{aligned}
\end{equation}
where $\mathbf{Q}^{p_k}_j,\mathbf{K}^{p_k}_j,\mathbf{V}^{p_k}_j$ represent query, key, and value for the $j$-th field in the $k$-th target permutation, respectively. $D$ denotes feature dimension of each feature. Here, query, key and value are transformed linearly from $\mathbf{E}^{p_k}_{;j}$, as follows:
\begin{equation}
  \begin{aligned}
    \mathbf{Q}^{p_k}_j \!=\! \mathbf{E}^{p_k}_{;j} \mathbf{W}^{Q}_j, \mathbf{K} \!=\! \mathbf{E}^{p_k}_{;j} \mathbf{W}^{K}_j, \mathbf{V} \!=\! \mathbf{E}^{p_k}_{;j}  \mathbf{W}^{V}_j,
    \forall j \!\in\! [N_f], \forall k \!\in\! [K], 
  \end{aligned}
\end{equation}
where $\mathbf{W}^{Q}_j, \mathbf{W}^{K}_j, \mathbf{W}^{V}_j \in \mathbb{R}^{D \times D}$. 
Then $\mathbf{H}^{p_k}_{j}$ are input into a  Multi-Layer Perceptrons (MLP) layer to generate representation:
\begin{equation}
  \begin{aligned}
    &  \mathbf{h}^{p_k}_j = \text{MLP}_1\Big(\mathbf{H}^{p_k}_{j} \Big), \ \  \forall j \!\in\! [N_f], \forall k \!\in\! [K], \\
    \mathbf{Z}^{p_k} = &\left[ \mathbf{h}^{p_k}_1 \ \ \  \mathbf{h}^{p_k}_2 \ \ \ \dots \ \ \ \mathbf{h}^{p_k}_{N_f}  \right] \in \mathbb{R}^{N_f \times D}, \ \ \forall k \in [K], 
  \end{aligned}
\end{equation}

After that, an inter-field self-attention layer is introduced to calculate the mutual influence between different fields in each permutation and output the final representation of the permutation $\mathbf{u}^{p}_k$ by an MLP layer, as follows:
\begin{equation}
  \begin{aligned}
    \mathbf{u}^{p}_k \!=\! \text{MLP}_2\Big(\mathbf{H}^{p_k}_{j} \Big) \!=\! \text{MLP}_2\Big( \text{soft} \max (\frac{\mathbf{Q'}_{\!\!p_k}{\mathbf{K'}_{\!\!p_k}}^\top}{\sqrt{D}})\mathbf{V'}_{\!\!p_k} \Big),  \forall k \!\!\in\!\! [K],
  \end{aligned}
\end{equation}
where $\mathbf{Q'}_{\!\!p_k},\mathbf{K'}_{\!\!p_k},\mathbf{V'}_{\!\!p_k}$ represent query, key, and value for the $k$-th permutation and are transformed linearly from $\mathbf{Z}^{p_k}$, as follows:
\begin{equation}
  \begin{aligned}
    \mathbf{Q'}_{\!\!p_k} = \mathbf{Z}^{p_k}\mathbf{W}^{Q'}, \mathbf{K'}_{\!\!p_k} = \mathbf{Z}^{p_k}\mathbf{W}^{K'}, \mathbf{V'}_{\!\!p_k} = \mathbf{Z}^{p_k}\mathbf{W}^{V'}, \forall k \!\in\! [K].
  \end{aligned}
\end{equation}

Analogously, the final representation of $m$-th permutation in user's permutation-level history click sequence is formulated as $\mathbf{u}^{b}_m$.
Through these two self-attention layers, we can 
effectively model the relationship between different items and different feature fields.

\subsubsection{Target Attention Unit (TAU)}
  \citet{zhou2018DIN} have proved that historical behaviors, which are more relevant to the target, can provide more information for the model's predict. We use a permutation-level target attention unit to model the interactions between target permutation and each permutation in historical behaviors, as follows:

 \begin{equation}
  \begin{aligned}
    &\mathbf{w}_{m;k} = \mathbf{u}^{b}_m \cdot \text{MLP}_{\text{Att}}\Big(\mathbf{u}^{p}_k||\mathbf{u}^{b}_m||(\mathbf{u}^{p}_k  \odot \mathbf{u}^{b}_m) ||(\mathbf{u}^{p}_k - \mathbf{u}^{b}_m) \Big), \\
    \mathbf{w}_{k} &= \text{Sum-Pool} \Big([\mathbf{w}_{1;k}, \dots, \mathbf{w}_{M;k}]
    \Big),   \forall m \in [M], \forall k \in [K].
  \end{aligned}
\end{equation}
where $||$ means concatenation. The output representation $\mathbf{w}_{k}$ of TAU is regarded as the representation of user's permutation-level interest on the target permutation and is used as input of next unit.

\subsubsection{Context-aware Prediction Unit (CPU)}

In Context-aware Prediction Unit, we use a parameter-sharing MLP layer to predict the list-wise pCTR of each item in each permutation.
Taking the $k$-th target permutation as an example, the inputs of CPU consist of four parts: representation of the $k$-th target permutation $\mathbf{u}^{p}_k$,
user's permutation-level interest on the $k$-th target permutation $\mathbf{w}_{k}$, 
the original predicted values (i.e., point-wise pCTRs of each item) $\{v^{p_k}_i\}_{i=1}^{N_d}$ and embedding matrix of the $k$-th target permutation $\mathbf{M}^{p}_k$. Then the list-wise pCTR of the $t$-th item in the $k$-th target permutation is predicted as follows:
\begin{equation}
  \begin{aligned}
    \hat{y}_{(k,t)} = \sigma\Big(\text{MLP}_3( \mathbf{u}^{p}_k ||\mathbf{w}_{k}||\{v^{p_k}_i\}_{i=1}^{N_d}||\mathbf{M}^{p}_k)\Big),
  \end{aligned}
\end{equation}
% wz-todo 下方cite里需加引文
where $\sigma$ is Sigmoid Function.
In PIER, the score of each permutation is easily obtained by summing the output list-wise pCTR. 
However, It should be noticed that by using the framework of PIER, the score of each permutation can be conveniently adjusted according to business needs, such as using GMV instead of CTR.

\subsection{Model Training}
\label{sec:mt}

In training stage, We first use the permutation of real exposures to train the OCPM, then jointly train the two modules by adding a contrastive learning loss.

\subsubsection{Pre-training of OCPM}

In order to accurately evaluate the selected top-K permutations, we first pre-train the OCPM using the samples collected from online log. The inputs are the permutations which are real displayed and the labels are whether the items in each displayed permutation are clicked. Then the loss of the OCPM is calculated as follows:
\begin{equation}
  \begin{aligned}
    Loss_{1} =  \sum_{t=1}^{N_d} \Big(- y_{(i,t)} \log (\hat y_{(k,t)})\! -\! (1\!-\! y_{(i,t)}) \log (1-\hat y_{(i,t)}) \Big),
  \end{aligned}
\end{equation}
where subscript $i$ is the index of samples and $t$ is the index of displayed items.
\subsubsection{Joint training of PIER}
\label{join_train}
In the joint training phase, since we use fix random vector as 'hash function' and 'position encoding', and the embeddings are shared between FPSM and OCPM, FPSM does not need to update gradients. When the embedding is updated, the hash signature is updated correspondingly. So how to improve the quality of the selected top-K permutations during the training process. Here, we propose a contrastive loss to ensure this goal. The main idea is that the difference between the average pCTR of the selected top-K permutations and the average pCTR of any K permutations in unselected permutations is as large as possible. Therefore, the final loss in the joint training phase is: 
\begin{equation}
  \begin{aligned}
    Loss_{2} =  -\sum_{k=1,k' \notin [K]}^K\Big(\frac{1}{N_d}\sum_{t=1}^{N_d}{\hat{y}_{(k,t)}} - \frac{1}{N_d}\sum_{t=1}^{N_d}{\hat{y}_{(k',t)}} \Big)^2,
  \end{aligned}
\end{equation}
where subscript $i$ is the index of samples and $t$ is the index of displayed items.

Finally, we sample a batch of samples $\mathcal{B}$ from the dataset and update PIER using gradient back-propagation w.r.t. the loss:
\begin{equation}
\label{eq:loss}
L(B) = \frac{1}{|\mathcal{B}|}\sum_{\mathcal{B}} \Big( Loss_{1} + \alpha \cdot Loss_{2}\Big),
\end{equation}
where $\alpha$ is the coefficient to balance the two losses.

\section{Experiments}
In this section, we conducted extensive offline experiments and online A/B test to evaluate the effectiveness of our framework\footnote{The code is publicly accessible at \url{https://github.com/Lemonace/PIER_code}.}. For offline experiments, we will first compare OCPM with existing baselines and analyze the role of different designs in it. Given the same prediction model, we will next compare the performance of our FPSM with other generative method. Finally, we will verify how different hyper-parameter settings (e.g., K, $\alpha$) affect the performance of our framework. For online experiments, we will compare our framework with the existing strategy deployed on the Meituan platform using an online A/B test.

\subsection{Experimental Settings}
\subsubsection{Dataset}

\begin{table}[bp]
  \caption{Statistics of the datasets. \ \ }
  \renewcommand\arraystretch{1.1}
  \centering
  \setlength{\tabcolsep}{3.5mm}{
  \begin{tabular}{c|ccc}
    \hline
  Dataset & \#Requests  &  \#Users  & \#Ads \\
  \hline
  \hline
  Avito & 53,562,269  &1,324,103   &  23,562,269 \\
  Meituan & 230,525,531  &3,201,922  & 98,525,531 \\
  \hline
  \end{tabular}
  }
  \label{tb:tb1}
\end{table}

In order to verify the effectiveness of our framework, we conduct sufficient experiments on both public dataset and industrial dataset. For public dataset, we choose Avito dataset. For industrial dataset, we use real-world data collected from Meituan food delivery platform. Table \ref{tb:tb1} gives a brief introduction of the datasets.

\noindent $\bullet$ \textbf{Avito}. The public Avito dataset contains user search logs and metadata from avito.ru, which contains more then 36M ads, 1.3M users and 53M search requests. The full features include user search information(e.g., userid, searchid,
and search date) and ad information(e.g., adid, categoryid, title
and so on). Each search id corresponds to a search page with multiple ads. For each user, We rank his search pages in increasing order based on the search date, and use the first T-1 search pages as the behavior pages, and the ads in the T-th search page as the target ads to be predicted. Here we use the data from 20150428 to 20150514 as
the training set and the data from 20150515 to 20150520 as the
testing set to avoid data leakage.

\noindent $\bullet$ \textbf{Meituan}. The industrial Meituan dataset is collected on Meituan food delivery platform during April 2022, which contains user information(e.g., userid, gender, age), ad information(e.g., adid, categoryid, brandid and so on). According to the date of data collection, we divide the
dataset into training and test sets with the proportion of 8:2.

\subsubsection{Evaluation Metrics}
For offline experiments, we use different metrics for different modules. Specifically, we use the widely adopted AUC metric to evaluate the effectiveness of the prediction module and use the following metrics to evaluate the whole framework:

\begin{itemize}[leftmargin=*]
  \item \textbf{HR(Hit Ratio)@1} \cite{alsini2020hit}. For each data, HR is 1 only when the top-K permutations selected by generative methods contains the best permutation.
  \item \textbf{Cost}. Cost means the overall time-consuming for different re-ranking frameworks.
\end{itemize}

For online experiments, we compare our proposed framework with existing method through CTR, GMV and inference time.

% TODO hyperparameters
\subsubsection{Hyperparameters}
The hidden layer sizes of $\text{MLP}_{\text{1}}$, $\text{MLP}_{\text{2}}$, 
$\text{MLP}_{\text{3}}$ are $(128, 64, 32)$, $(60, 32, 20)$, and $(50, 20)$, respectively.  the learning rate is $10^{-3}$, the optimizer is Adam and the batch size is 1,024. The $\alpha$ is 0.1, the embedding size is 8 and the length of user behavior sequence is 5. For Avito dataset, the length of ranking list and re-ranking list are both 5, thus the length of full permutation is 120 and $\textbf{K}$ is set to 10. For Metuan dataset page, we select 3 items from the initial ranking list which contains 10 items, thus the length of full permutation is 720 and $\textbf{K}$ is set to 100. 

% %% 实验结果表

\begin{table}[]

  \caption{The experimental results about AUC and Logloss on two datasets.}
  \renewcommand\arraystretch{1.1}
  \centering
  \setlength{\tabcolsep}{2.7mm}{
  \begin{tabular}{c|cc|cc}
    \hline
  \multirow{2}{*}{Model} & \multicolumn{2}{c|}{Avito} & \multicolumn{2}{c}{Meituan} \\
                         & AUC         & LogLoss     & AUC          & LogLoss      \\
                     
  \hline
  \hline
  DNN                    & 0.6876      & 0.0483      & 0.6538       & 0.1917       \\
  DCN                    & 0.6896      & 0.0483      & 0.6550       & 0.1916       \\
  PRM                    & 0.7131      & 0.0481      & 0.6718       & 0.1899       \\
  EXTR                   & 0.7114      & 0.0481      & 0.6704       & 0.1901       \\
  Edge-Rerank            & 0.7163      & 0.0479      & 0.6694       & 0.1909       \\
  OCPM                   & \textbf{0.7320 }      & \textbf{0.0471}      & \textbf{0.6822}       & \textbf{0.1891} \\
  \hline  
 
  \end{tabular}
  }
  \label{tb:tb2}
  \end{table}

\subsection{Offline Experiments For OCPM}
\subsubsection{Baselines}
We compare OCPM with both point-wise and list-wise representative methods as baselines. We select DNN, DCN as point-wise baselines and PRM, EXTR and Edge-Rerank from Kuaishou as list-wise baselines. A brief introduction of these methods are as follows:
\begin{itemize}[leftmargin=*]
  \item \textbf{DNN} \cite{covington2016dnn}. DNN is a basic deep learning method for CTR prediction, which applies MLP for high-order feature interaction.
  
  \item \textbf{DCN} \cite{wang2017dcn}. 
  DCN explicitly applies feature crossing at each layer, requires no manual feature engineering, and adds negligible extra complexity to the DNN model. 
  
  \item \textbf{PRM} \cite{pei2019prm}. PRM adjusts an initial list by appling the self-attention mechanism to capture the mutual influence between items.
  
 \item \textbf{EXTR} \cite{chen2022extr}. EXternality TRansformer regards target ad with all slots as query and external items as key\&value to model externalities in all exposure situations.
 
  \item \textbf{Edge-Rerank}  \cite{gong2022real}. Edge-Rerank combines an on-device ranking model and an adaptive beam search method to generate context-aware re-ranking result.
\end{itemize}

\subsubsection{Performance Comparison}

Table \ref{tb:tb2} summarizes the results of offline experiments. All experiments were repeated 5 times and the averaged results are reported. We have the following observations from the experimental results: i) All re-ranking listwise model (e.g. PRM, EXTR) makes great improvements over point-wise model (e.g. DNN, DCN) by modeling the mutual influence among contextual items, which verifies the impact of context on user clicks behavior. ii) Compare with transformer-based model(e.g. PRM, ETXR), Edge-Rerank also improve the CTR prediction because of they use the history sequence and previous item information. iii) Our proposed OCPM brings 0.0189/0.0104 absolute AUC on Avito/Metuan dataset gains over the state-of-the-art independent baseline which is a significant improvement in industrial recommendation system.

\subsubsection{Ablation Study}
To explore the effectiveness of different modules in OCPM, we conduct ablation studies on Avito and Meituan dataset. All experiments were repeated 5 times and the averaged AUC is reported:
\begin{itemize}[leftmargin=*]
  \item OCPM (-OAU) blocks the omnidirectional attention module. OAU aims to capture item context and feature context information, which could model the competitive relationship between the same attribute of different item. As shown in Table \ref{tb:tb3}, AUC decreases by 0.0112/0.0048, suggesting that extracting context information is crucial and the proposed OAU meets this requirement.
  \item  OCPM (-TAU) does not use the target attention unit. TAU aims to compute the interactions of target permutation and each permutation in historical behavior. As shown in Table \ref{tb:tb3}, AUC decreases by 0.0057/0.0036 on Avito/Metuan dataset, suggesting that our TAU are capable to capture user history page-level interest.
\end{itemize}

  \begin{table}[tb]
    \caption{Result of ablation experiment on different parts in re-ranking model.}
    \renewcommand\arraystretch{1.1}
    \centering
    \setlength{\tabcolsep}{3.4mm}{
    \begin{tabular}{l|cc|cc}
      \hline
    \multirow{2}{*}{Model} & \multicolumn{2}{c|}{Avito} & \multicolumn{2}{c}{Meituan} \\
                           & AUC         &   LogLoss   & AUC          & LogLoss      \\
                       
    \hline
    \hline
    OCPM                    & \textbf{0.7320}      & 0.0483      & \textbf{0.6822}       & 0.1917       \\
    \quad - OAU             & {0.7198}      & 0.0483      & 0.6774       & 0.1916       \\
    \quad - TAU             & {0.7263}      & 0.0481      & 0.6786       & 0.1899       \\
    \hline  
    \end{tabular}
    }
    \label{tb:tb3}
    \end{table}

\subsection{Offline Experiments For PIER}

\subsubsection{Baselines}
We compare our whole framework which combines FPSM and OCPM with some heuristic generative methods as baselines. All these methods use a fixed OCPM but the candidate permutations are generated in different ways. A brief introduction of these methods is as follows:
\begin{itemize}[leftmargin=*]
  \item \textbf{Random \& OCPM}. We randomly select K permutations as candidates. 
  
  \item \textbf{PRS(Beam Search \& OCPM)}  \cite{feng2021PRS}. 
 We use the beam search method to generate K candidate permutations based on the cumulative CTR.
  
  \item \textbf{Full Permutation \& OCPM}. 
We directly feed all candidate permutations into the prediction model and select K permutations based on average CTR. This can be seen as the upper bound of our framework.
\end{itemize}

\subsubsection{Performance Comparison}

Table \ref{tab:4} summarizes the results of offline experiments and we have the following observations: i) Intuitively, PIER achieves great improvements over random and beam-search methods on finding the best permutation on both public dataset and industrial dataset with a small increase in time complexity. One reasonable explanation is that FPSM can select better permutations through the guidance of contrastive loss. ii) We still have much room for improvement compared with the full permutation method on finding the best permutation, but we greatly reduce the time cost of the whole framework.

\subsubsection{Ablation Study}
To verify the impact of different units (i.e., time-aware weighting, contrastive loss), we study two ablated variants of PIER framework:
\begin{itemize}[leftmargin=*]
  \item PIER (-time-aware weighting) does not use the time-aware weight on each behavior and treat them as equally important.
  \item PIER (-contrastive Loss) removes the contrastive loss for improving the quality of the selected top-K permutations, i.e., $\alpha=0$.
\end{itemize}

Since removing a certain unit requires re-training the entire framework, we try to ensure the AUC of OCPM are as close as possible to guarantee the comparability of the ablation experiment. Judging from the experimental results, we have the following findings: 
i) The performance gap between w/ and w/o time-aware weighting on distance calculation has little impact on public dataset but the performance on industrial dataset is affected to some extent. This indicates that in the Meituan scenario, users' interests remain consistent in a short period of time.
ii) The decline in the performance without contrastive loss is obvious on both dataset. This indicates that the auxiliary loss enables FPSM to select better permutations.

    \begin{table}[tb]
      \caption{HR and Cost of different methods.}
      \renewcommand\arraystretch{1.1}
      \centering
      \setlength{\tabcolsep}{1.2mm}{
    \begin{tabular}{l|cc|cc}
    \hline
    \multirow{2}{*}{     \ \ \ \ \ \ \ \  \ \ \ \ \ \ \ \ \  Model}       & \multicolumn{2}{c|}{Avito}                                       & \multicolumn{2}{c}{Meituan}                                      \\
                                 & HR &  Cost (ms) & HR &  Cost (ms)\\ \hline \hline
    Random \& OCPM                & 0.09  & \textbf{10.8}                                                           & {0.13}  & \textbf{70.5}                                                           \\
    PRS (Beam-Search \& OCPM)      & 0.62  & {16.8}                                                           & 0.72  & 78.7                                                           \\
    Full Permutation \& OCPM      & 1.00  &  100.5                                                          & 1.00  & 430.5                                                           \\\hline   \hline    
    PIER (FPSM \& OCPM)            & \textbf{0.78}  & {17.5}                                                           & \textbf{0.84}  & {85.3}                                                        \\ 
    \quad - time-aware weighting & 0.74  & 17.3                                                           & 0.70  & 84.8                                                          \\
    \quad - contrastive loss    & 0.52  & 17.5                                                           & 0.61  & 85.1                                                            \\ \hline
    \end{tabular}
      }
      \label{tab:4}
    \end{table}
    
\begin{table}[hbtb]
  \caption{Result of ablation experiment on different parts in PIER}
  \renewcommand\arraystretch{1.1}
  \centering
  \setlength{\tabcolsep}{1.7mm}{
\begin{tabular}{l|ccc|ccc}
\hline
\multirow{2}{*}{Settings}       & \multicolumn{3}{c|}{Avito}    & \multicolumn{3}{c}{Meituan}     
\\
& AUC & HR &  Cost (ms)  & AUC & HR &  Cost (ms)  \\ \hline \hline
$\alpha=0$               & 0.7337    & 0.52  & 17.5             & 0.6833    & 0.57  & 85.1                                                            \\
$\alpha=0.01$      &  0.7336   & 0.63  & 17.4                     & 0.6831    & 0.62  & 85.3                                                      \\
$\alpha=0.05$      & 0.7329    & 0.71  & 17.6                       & 0.6828    & 0.71  & 85.3                                               \\
$\alpha=0.1$            & 0.7320   & 0.78  & 17.5                 & 0.6822    & 0.84  & 85.2                                                      \\ 
$\alpha=0.3$            & 0.7108    & 0.81  & 17.4                   & 0.6674    & 0.88  & 85.4                                                        \\
$\alpha=0.5$            &  0.6927  & 0.88  & 17.6                    & 0.6425    & 0.91  & 85.3                                                     \\
\hline  \hline
$K=5$ & 0.7325    & 0.63  & 13.8                             & -   & -  & -                                   \\
$K=10$ & 0.7320    & 0.71  & 17.5                              & -   & -  & -                                    \\
$K=20$  & 0.7318   & 0.82  & 25.2                             & -   & -  & -                                    \\
$K=50$ & 0.7299    & 0.93   & 46.9                            & 0.6827    & 0.78  & 49.2                                    \\
$K=100$ & -   & -  & -                             & 0.6822    & 0.84  & 85.3                                       \\   
$K=200$ & -   & -  & -                                & 0.6813    & 0.92  & 157.3                                    \\
$K=300$ & -   & -  & -                                & 0.6799    & 0.95  & 223.3                                    \\\hline
\end{tabular}
  }
  \label{label:alpha}
\end{table}

\subsection{Hyperparameter Analysis}
  We analyze the sensitivity of two hyperparameters: $\alpha$, $K$. 
  Specifically, $\alpha$ is the weight of the contrastive loss  and $K$ is number of permutations selected by FPSM.
  The result is shown in table \ref{label:alpha}, showing the same trend on public dataset and industrial dataset and we have the following findings :  
  
  i) As $\alpha$ increases within a certain range, the AUC of OCPM maintains a relatively good level, while HR is improved. When $\alpha$ exceeds a certain level, the influence of the contrastive loss on OCPM increases, and the AUC begins to decline, resulting in a decrease in the confidence of the HR indicator.
  
  ii) Changing $K$ mainly affects HR and average cost of the framework. Increasing $K$ within a certain range can quickly improve the HR performance. When it exceeds a certain range, HR increases slowly. Meanwhile, the average cost increases with the increase of $K$. Therefore, the value of $K$ should be set reasonably to balance the effect and efficiency in practice.

        \begin{table}[]
      \caption{The experimental results from Online A/B testing.}
      \renewcommand\arraystretch{1.15}
      \centering
      \setlength{\tabcolsep}{0.9mm}{
        \begin{tabular}{l|ccccc}
          \hline
        Model  & CTR & GMV & Cost(ms) & Time-out\\
        \hline
        \hline
        Base Ranking Model &  0\% & 0\%  & \textbf{0} & 0.0\% \\
        PRM &   1.21\% & 0.92\%  & 6.1 & 0.0\% \\
        Beam-Search \& Evaluator &  3.17\% & 2.54\%  & 10.5 & 0.052\% \\
        Full Permutation \& Evaluator&  - & -  & 52.4 & 64.39\% \\
        \textbf{PIER} &  \textbf{5.46\%} & \textbf{5.83\%}  & {11.1} & \textbf{0.061\%} \\
        \hline
        
        \end{tabular}
      }
      \label{tb:tb6}
    \end{table}

\subsection{Online Results}
  We compare PIER with other models (e.g. base ranking model, PRM and so on) and all deployed on Meituan food delivery platform through online A/B test. Specifically, we conduct online A/B test with 1\% of whole production traffic from April 09, 2022 to April 25, 2022 (one week). As a result, we find that PIER gets CTR and GMV increase by 5.46\% and  5.83\% respectively. Besides, we focus on time costs, which is an important indicator, determine whether it can be applied to a large scale of industrial scenarios. As show in Tabel \ref{tb:tb6}, Full-Permutation could not deploy because of time-out ratio increase by 64.39\%. Compared with beam-search, PIER has improved CTR by 2.29\% and GMV by 2.31\% while the time-out ratio effect increases little, which is acceptable to the system. Now, PIER has been deployed online and serves the main traffic, and contributes to significant business growth.

\section{Conclusions}

This paper presents a novel end-to-end re-ranking framework
named PIER which contains two mainly modules named
FPSM and OCPM. Inspired by long-time user behavior modeling
methods, we apply SimHash in FPSM to select top-K candidates
from the full permutation. For better capturing the context information in the permutation, we design a novel omnidirectional attention mechanism in OCPM. Finally, we jointly train these two modules in an end-to-end way by introducing a comparative learning loss to guide the FPSM to generate better
permutations. Both offline experiment and online A/B test show that PIER significantly outperformed other existing re-ranking baselines, and we have deployed PIER on Meituan food delivery platform.

\newpage
\balance
\bibliographystyle{ACM-Reference-Format}
\bibliography{pier}

\end{document}